\documentclass{article}

\usepackage[english]{babel}
\usepackage{bm}
\usepackage{amsmath, amsthm, amssymb,amsbsy,amsfonts}
\usepackage{verbatim}
\usepackage{bbold}
\usepackage{bbm}
\usepackage{mathbbol}
\usepackage{graphicx,epsfig,color}
\usepackage{cite}
\usepackage{tikz}

\def\cO{{\cal O}}
\def\vev#1{{\left\langle #1 \right\rangle}}

\title{Holographic impurities and Kondo effect}
\author{ 
J.~Erdmenger${}^1$, M.~Flory${}^1$, C.~Hoyos${}^2$,\\ M-N.~Newrzella${}^1$, A.~O'Bannon${}^3$, J.~Wu${}^4$,\\
\small \em ${}^1$ Max-Planck-Institut f\"ur Physik (Werner-Heisenberg-Institut),\\
\small \em F\"ohringer Ring 6, D-80805 Munich, Germany.\\
\small \em ${}^2$ Department of Physics, Universidad de Oviedo,\\
\small \em Avda. Calvo Sotelo 18, 33007, Oviedo, Spain.\\
\small \em ${}^3$ STAG Research Centre, Physics and Astronomy, University of Southampton,\\
\small \em Highfield, Southampton SO17 1BJ, United Kingdom.\\
\small \em ${}^4$ Department of Physics and Astronomy, University of Alabama,\\
\small \em Tuscaloosa, AL 35487, USA.
}

\begin{document}
\maketitle

\begin{abstract}
Magnetic impurities are responsible for many interesting phenomena in condensed matter systems, notably the Kondo effect and quantum phase transitions. Here we present a holographic model of a magnetic impurity that captures the main physical properties of the large-spin Kondo effect. We estimate the screening length of the Kondo cloud that forms around the impurity from a calculation of entanglement entropy and show that our results are consistent with the $g$-theorem. 
 \end{abstract}
 
\section{Introduction}

\begin{figure}
\begin{center}
\includegraphics[width=8cm]{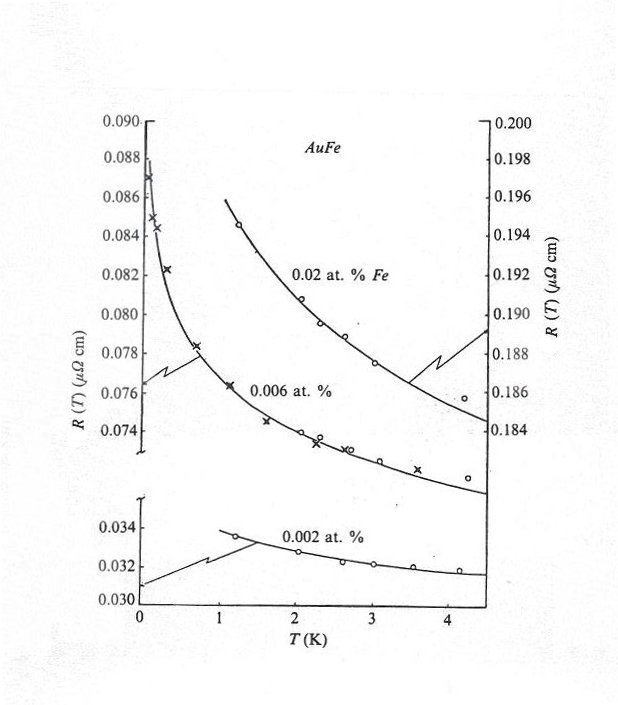}
\caption{\small \em Raise of the resistivity at low temperatures. Points and crosses are experimental results for iron impurities in gold, while lines are theoretical predictions (figure from \cite{Kondo01071964}).} \label{fig:kondo}
\end{center}
\end{figure}

Magnetic impurities have local magnetic moment due to the spin of unpaired electrons. Their interactions with the electrons in a conducting material lead to interesting phenomena that affect to the physical properties the material.  The best known is the Kondo effect, a raise in the resistivity at low temperatures first explained by Jun Kondo in 1964 \cite{Kondo01071964} (see Figure \ref{fig:kondo}).  The Kondo effect has been observed in many systems, the canonical examples being metals doped with magnetic iron impurities \cite{1974RPPh...37..147R,kondobook1} and quantum dots \cite{1998Natur.391..156G,1998Sci...281..540C,2000Sci...289.2105V}. Kondo used a simple model of a Landau Fermi liquid (LFL) of electrons interacting with a localized spin, which was seminal to Wilson's development of Renormalization Group (RG) techniques~\cite{Wilson:1974mb,PhysRevB.21.1003,PhysRevB.21.1044}, integrability~\cite{PhysRevLett.45.379,Wiegmann:1980,RevModPhys.55.331,
doi:10.1080/00018738300101581,0022-3719-19-17-017,1994cond.mat..8101A,
ZinnJustin1998, 
PhysRevB.58.3814}, large-$N$ limits~\cite{PhysRevB.35.5072,RevModPhys.59.845,1997PhRvL..79.4665P,
1998PhRvB..58.3794P,2006cond.mat.12006C,2015arXiv150905769C}, Conformal Field Theories (CFTs)~\cite{1998PhRvB..58.3794P,Affleck:1990zd,Affleck:1990by,
Affleck:1990iv,Affleck:1991tk,
PhysRevB.48.7297,Affleck:1995ge} and more~\cite{Hewson:1993,doi:10.1080/000187398243500}. 

The interaction between the magnetic impurity and the electrons is characterized by a dynamically generated scale, conventionally expressed as the Kondo temperature $T_K$. At higher temperatures the magnitude of the coupling decreases, thus providing an early example of asymptotic freedom. At temperatures below $T_K$ the impurity is screened by the electrons, eventually leading to an infrared (IR) fixed point as the temperature goes to zero. 

\begin{figure}
\begin{center}
\includegraphics[width=6cm]{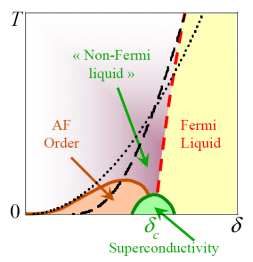}
\caption{\small \em A cartoon of the phase diagram in heavy fermion compounds. As the parameter $\delta$ is varied at zero temperature, there is a transition from an antiferromagnetic state (AF) to a Fermi liquid of heavy fermions. The quantum critical point is at $\delta=\delta_c$, but in this case it is depicted cloaked by a superconducting dome. At low temperatures around the critical point, the system behaves like a Non-Fermi liquid.  The dotted and dashed lines indicate the temperature for the transition due to RKKY interactions and the Kondo temperature, respectively (figure from \cite{phaseHF} ).}\label{fig:phaseHF}
\end{center}
\end{figure}

When many impurities are present the effective interactions among them can be important. For a lattice of impurities these interactions are of Ruderman-Kittel-Kasuya-Yosida (RKKY) type, meaning that they are mediated by the conduction electrons when they scatter with different impurities, and are ferromagnetic or anti-ferromagnetic in nature. They thus tend to drive the system to a ferromagnetic or anti-ferromagnetic ground state, while the Kondo effect tends to screen the spins and produce a non-magnetic ground state. The competition between these two kinds of interactions can induce a quantum phase transition \cite{HFColeman2006,HFGegenwart2008,HFSi2010,2010uqpt.book..193S,2010Sci...329.1161S,2015arXiv150905769C} as parameters of the system, such as impurity concentration or pressure, are varied. Non-Fermi liquid behavior and unconventional superconductivity has been associated in heavy fermion compounds to the quantum critical behavior close to the transition. On the other hand the large mass renormalization in the Fermi liquid region has been attributed to the Kondo effect (see Figure~\ref{fig:phaseHF}). Although the qualitative picture is understood, the full derivation of the phase diagram from the impurity model is as yet an open problem.

A distinct possibility is that close to the quantum critical point the electrons interacting with the impurities cannot be treated as a LFL, so it would be interesting to have a model of magnetic impurities coupled to strongly correlated electrons. It would also be interesting for other cases where the electrons are in a non-Fermi liquid state, such as a Luttinger liquid in 1+1 dimensions. Both perturbative and Monte Carlo methods are not well suited for this kind of problem, and exploring alternative approaches seems a worthwhile enterprise. A natural proposal is to use holographic models, as formulated in the AdS/CFT correspondence~\cite{Maldacena:1997re,Gubser:1998bc,Witten:1998qj}. In this context a handful of models of impurities have been constructed~\cite{Kachru:2009xf,
Kachru:2010dk,Faraggi:2011bb,Jensen:2011su,Karaiskos:2011kf,Harrison:2011fs,Benincasa:2011zu,Benincasa:2012wu,Faraggi:2011ge,Itsios:2012ev,Erdmenger:2013dpa}, but in most cases they restrict to the physics of the conformal fixed point where the impurity is seen as a 0+1-dimensional conformal defect. The exception is the model of \cite{Erdmenger:2013dpa}, that we will present here for a single impurity.  This model realizes the RG flow of the original Kondo model and can be put at nonzero temperature.  In the following we will present it and discuss how it captures the main aspects of the Kondo effect. We will also show that the contribution to entanglement entropy of the impurity and the $g$-theorem have very simple geometric realizations in the holographic model. We should mention that the model has been generalized to two impurities in \cite{O'Bannon:2015gwa}, where also the competition between RKKY and Kondo interactions is studied.

\section{Impurities in field theory and large-$N$}\label{sec:fieldth}

The interaction between electrons and magnetic impurities can be described as the coupling of the electron spin $\vec{J}=\psi^\dagger \frac{\vec{\tau}}{2}\psi$  \footnote{$\vec{\tau}$ are the Pauli matrices.} and the impurity spin $\vec{S}$. The effective Hamiltonian is
\begin{equation}\label{eq:ham}
H=\psi^\dagger \frac{-\nabla^2}{2m}\psi+\lambda_K \delta^{(3)}(x)\, \vec{S}\cdot \vec{J}.
\end{equation}
Where $\lambda_K<0$ is the anti-ferromagnetic Kondo coupling. Only the $s$-wave component of the electrons interacts with the impurity, so the model can be reduced to a chiral theory in $1+1$ dimensions coupled to a 0+1-dimensional defect . In this description it is easy to see that the Kondo coupling is marginally relevant, so it vanishes in the UV and grows in the IR until it reaches a fixed point. The properties of the IR CFT can be inferred from spectral flow arguments  \cite{1998PhRvB..58.3794P,Affleck:1990zd,Affleck:1990by,
Affleck:1990iv,Affleck:1991tk,PhysRevB.48.7297,Affleck:1995ge}. If the impurity is placed directly in a $1+1$ dimensional theory, the situation is similar but instead of a LFL the effective description of the electrons should be a Luttinger liquid.

In \eqref{eq:ham} there is an explicit $SU(2)$ symmetry of rotations of the spin of the electrons, and a $U(1)$ particle number symmetry that changes the phase. For non-relativistic fermions of spin $s$ the symmetry is $SU(N)$ ($N=2s+1$), and if there are $k$ channels interacting with the impurity (as may happen in quantum dots), the particle number symmetry is extended to $SU(k)\times U(1)$ . Using the $1+1$ chiral description it can be shown that the symmetry is actually $SU(N)_k\times SU(k)_N\times U(1)_{kN}$. 

The spin of the impurity can be in different representations of the spin group, each associated to a Young tableau. For a totally antisymmetric representation, corresponding to a column with $q<N$ boxes, it will be convenient to rewrite the spin operator in terms of slave fermions $\chi$ in the fundamental representation of $SU(N)$ \footnote{Other representations can be realized by having  more than one `flavor' of slave fermions.} \cite{PhysRevB.58.3794,PhysRevLett.90.216403}
\begin{equation}
S^a=\chi^\dagger T^a \chi,
\end{equation}
where $T^a$ are the $SU(N)$ generators. Note that there is an additional $U(1)$ symmetry that rotates the phase of the fermions $\chi$. Since we are introducing additional degrees of freedom we need to impose a constraint to restrict to the subspace of physical states. This amounts to fixing the charge to the number of boxes in the Young tableau
\begin{equation}\label{eq:qcons}
\chi^\dagger \chi=q.
\end{equation}
Fierz identities allow to write the Kondo coupling as a double trace coupling for a scalar operator $\cO=\psi^\dagger\chi$
\begin{equation}\label{eq:kondocoup}
\lambda_K  \vec{S}\cdot \vec{J} \simeq \frac{1}{2}\lambda_K\left[ \cO \cO-\frac{q}{N}\psi^\dagger\psi\right].
\end{equation}
In the large-$N$ limit the system can be solved using a mean field approximation. There is a phase transition such that the scalar operator condenses below a critical temperature $T_c$, $\vev{\cO}\neq 0$ for $T\leq T_c$ \cite{PhysRevB.58.3794,PhysRevLett.90.216403}. This is interpreted as the formation of a bound state between the impurity and the conduction electrons, i.e. the screening of the impurity. Spontaneous symmetry breaking is possible because of the large-$N$ limit, but fluctuations of the order parameter grow in time at finite $N$ and would eventually wash away the phase transition. However, this would happen only at times that grow exponentially with $N$.

The holographic model is inspired by the holographic dual to a configuration of D3, D5 and D7 branes (see \cite{Erdmenger:2013dpa} for details) that contain all the main ingredients of the construction above: chiral fermions in $1+1$ dimensions coupled to slave fermions in a $0+1$-dimensional defect. There is an important difference, though. In the D-brane construction the spin group is gauged and in order to have a reliable supergravity approximation, the coupling of the spin group has to be very large. Therefore, the impurity is not coupled to a LFL but rather to strongly correlated fermions and dynamical gauge fields. 

\section{Holographic model of the Kondo effect}\label{sec:model}

\begin{figure}
\begin{center}
\includegraphics[width=8cm]{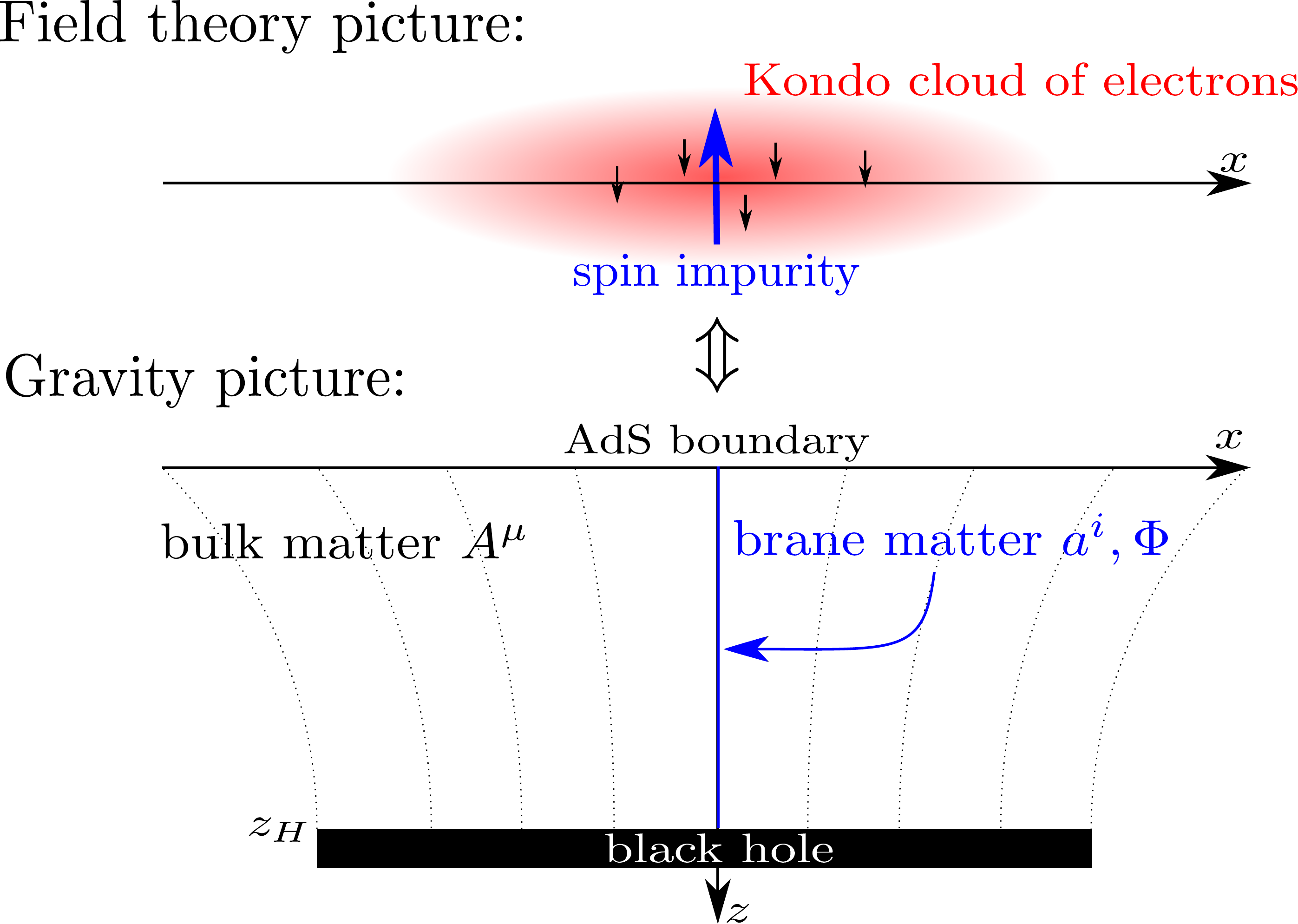}
\caption{\small \em Holographic model of a magnetic impurity. In the gravity side the impurity is a defect extending between the boundary and the horizon. The spin current, the charge of the slave fermions and the operator $\cO$ are mapped to gauge fields and a scalar field respectively (Figure from \cite{Erdmenger:2015spo}).}\label{fig:holoK}
\end{center}
\end{figure}

Figure~\ref{fig:holoK} summarizes the holographic construction of the large-$N$ impurity model. We will restrict to a single channel and a completely antisymmetric representation of the impurity spin.

In the UV the theory is a $1+1$ CFT, so the holographic dual is asymptotically $AdS_3$, actually a BTZ geometry if the system is at finite temperature  $T=1/(2\pi z_H)$
\begin{equation}
\begin{split}
ds^2=& g_{\mu\nu}dx^\mu dx^\nu=\frac{L^2}{z^2}\left(\frac{dz^2}{f(z)}-f(z)dt^2+dx^2\right), \\ &f(z)=1-z^2/z_H^2.
\end{split}
\end{equation}
In the bulk geometry there is a gauge field $A_\mu$ dual to the $U(1)_N$ particle number current of the fermions
\begin{equation}
S_{A}=\frac{N}{4\pi}\int  A \wedge d A.
\end{equation}
The coefficient $N$ fixes the rank of the spin group. The spin group does not appear explicitly in the holographic construction because it is gauged. The impurity is represented by a $1+1$-dimensional defect inside the geometry that extends from the position of the impurity at the $AdS$ boundary to the horizon. On the defect there is a scalar field $\Phi$ dual to the operator $\cO$ and a $U(1)$ gauge field $a_m$ dual to the $U(1)$ charge of the slave fermions. The action is
\begin{equation}
S_{\rm imp}=-{\cal N}\int_{AdS_3} \delta(x) \left(\frac{1}{4}f_{mn} f^{mn} +(D\Phi)^2+V(\Phi^\dagger \Phi)\right).
\end{equation}
We will use greek indices for the $AdS_3$ directions and latin indices for the directions along the defect. ${\cal N}$ is a normalization factor, $f_{mn}$ is the field strength of $a_m$ and the covariant derivative acting on the scalar field is
\begin{equation}
D_m\Phi=\partial_m\Phi-ie a_m \Phi+ie A_m\Phi.
\end{equation}
Note that large-$N$ scaling is ${\cal N}\sim 1/e\sim N$.

\subsection{Impurity spin and Kondo coupling}

The representation of the impurity is determined by the constraint \eqref{eq:qcons}. The charge $q$ of the slave fermions is determined by the leading coefficient $Q$ in the asymptotic expansion of the gauge field as $z\to 0$ \footnote{More precisely $q$ is proportional to the flux ${\cal C}$, with  ${\cal C}^2=-\frac{1}{2}f_{mn}f^{mn}$, but one can determine ${\cal C}$ from $Q$.}
\begin{equation}
a_t\simeq \frac{Q}{z}+\mu.
\end{equation}
The subleading term $\mu$ is the chemical potential on the impurity. By fixing $Q$ we fix the representation, note that within the large-$N$ approximation $Q$ is a continuous parameter.  The potential in the construction is arbitrary, for simplicity we will make it quadratic
\begin{equation}
V(\Phi^\dagger\Phi)=M^2\Phi^\dagger\Phi.
\end{equation}
By imposing $M^2=Q^2-\frac{1}{4}$,
we fix the dimension of the scalar operator $\cO$ to be $\Delta=1/2$ in the UV. This is necessary in order to have a marginal Kondo coupling  \eqref{eq:kondocoup}.

The leading coefficients in the asymptotic expansion of the scalar field are
\begin{equation}
\Phi(z)\simeq \alpha_\Lambda \sqrt{z}\log(\Lambda z)+\beta_\Lambda \sqrt{z},
\end{equation}
where $\alpha_\Lambda$ is dual to a source for the $\cO$ operator, while $\beta_\Lambda$ is dual to the expectation value (vev).  Following  \cite{Witten:2001ua}, a double trace coupling translates to a relation between the leading coefficients in the expansion of the scalar field\footnote{The double trace deformation  changes the source $\alpha_\Lambda \to \alpha_\Lambda-\kappa_\Lambda\beta_\Lambda$.}
\begin{equation}
\alpha_\Lambda=\kappa_\Lambda\beta_\Lambda,
\end{equation}
where $\kappa_\Lambda\propto \lambda_K$ determines the Kondo coupling at the scale $\Lambda$.  The value of the scalar field $\Phi(z)$ should be independent of the scale $\Lambda$. This leads to the following value of the Kondo coupling at the scale $2\pi T$
\begin{equation}
\kappa_T=\frac{\kappa_\Lambda}{1+\kappa_\Lambda \log\left(\frac{\Lambda}{2\pi T} \right)}.
\end{equation}
For an anti-ferromagnetic coupling $\kappa_\Lambda <0$ we recover asymptotic freedom, since  $\kappa_T \to 0^-$ as $T\to \infty$. The coupling diverges at the Kondo temperature $T_K=\frac{\Lambda}{2\pi} e^{1/\kappa_\Lambda}= T e^{1/\kappa_T}$.  Below the Kondo temperature $T<T_K$, the effective coupling becomes ferromagnetic $\kappa_T>0$.

\subsection{Phase transition}

There are two possible phases characterized by the scalar field. When $\Phi=0$ the vev of the dual operator  vanishes $\vev{\cO}=0$ and we are in the normal phase. On the other hand, if the scalar field has a non-trivial profile $\Phi(z)$, the dual operator acquires an expectation value $\vev{\cO}\neq 0$ that we have identified with the screening in the impurity, we will call this the condensed phase. Which phase is the true ground state of the system can be determined by comparing their free energies and picking the one of minimal value. The free energy can be computed in the holographic model in the usual way, by evaluating the on-shell action properly renormalized. 

\begin{figure}
\begin{center}
\includegraphics[width=8cm]{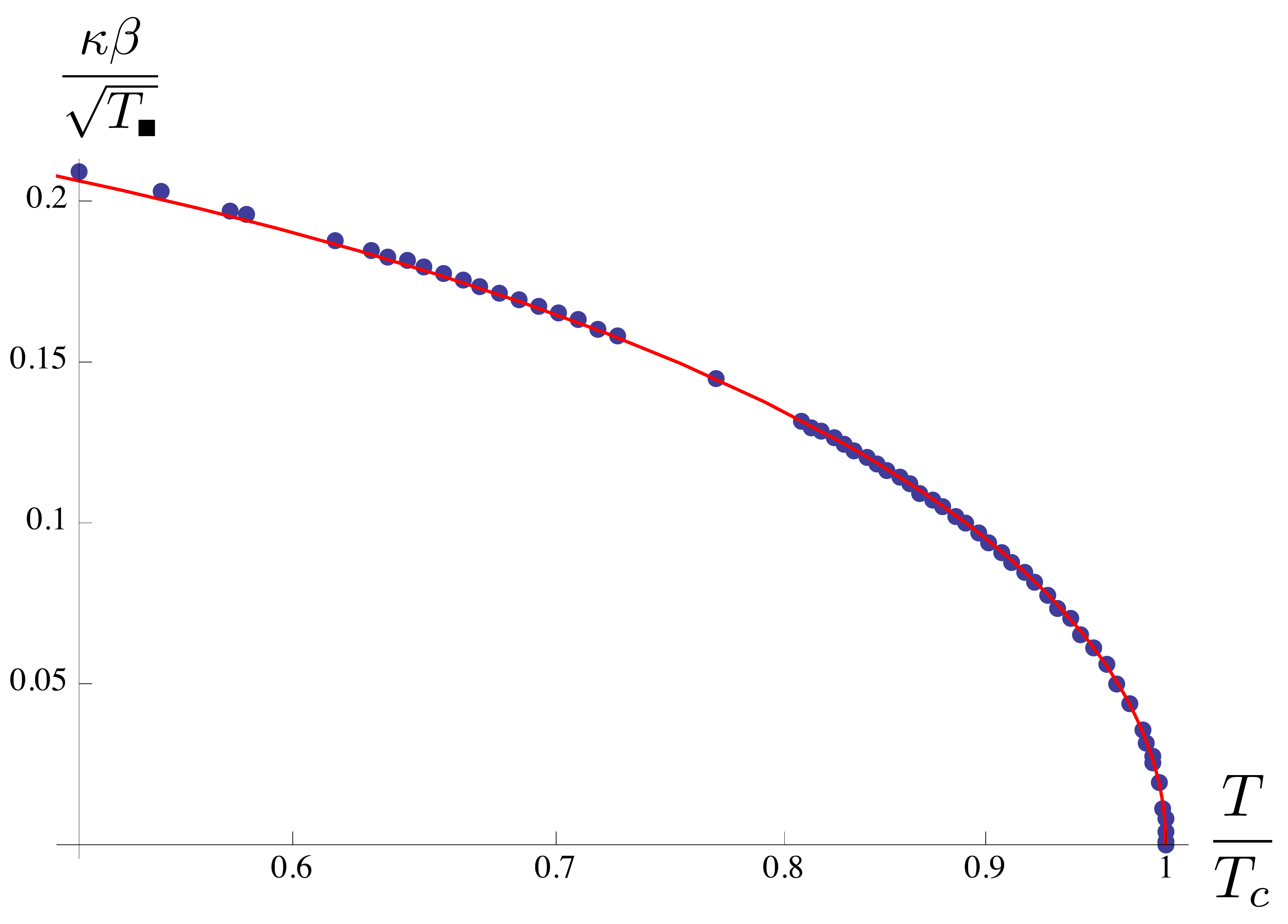}
\caption{\small \em For $T<T_c$ the the vev of the operator $\cO$ is nonzero. The solid red line fits numerical data to $0.3(1-T/T_c)^{1/2}$. The phase transition is second order and of mean field type (Figure from \cite{Erdmenger:2013dpa}).}\label{fig:phastrans}
\end{center}
\end{figure}

The result of this thermodynamic analysis is that at high temperatures the normal phase is favored, but there is a critical temperature $T_c\sim 0.9 T_K$ at which the free energy of the condensed phase starts to be lower. In fact, for temperatures below $T_c$ the normal phase becomes unstable. The transition between the two phases is second order and of mean field type. The vev $\vev{\cO}\sim \beta$ grows at lower temperatures as depicted in Figure~\ref{fig:phastrans}. The holographic model thus exhibits the same physics as the large-$N$ field theory model of an impurity coupled to a LFL.

\subsection{Backreaction}

\begin{figure}
\begin{center}
\includegraphics[width=8cm]{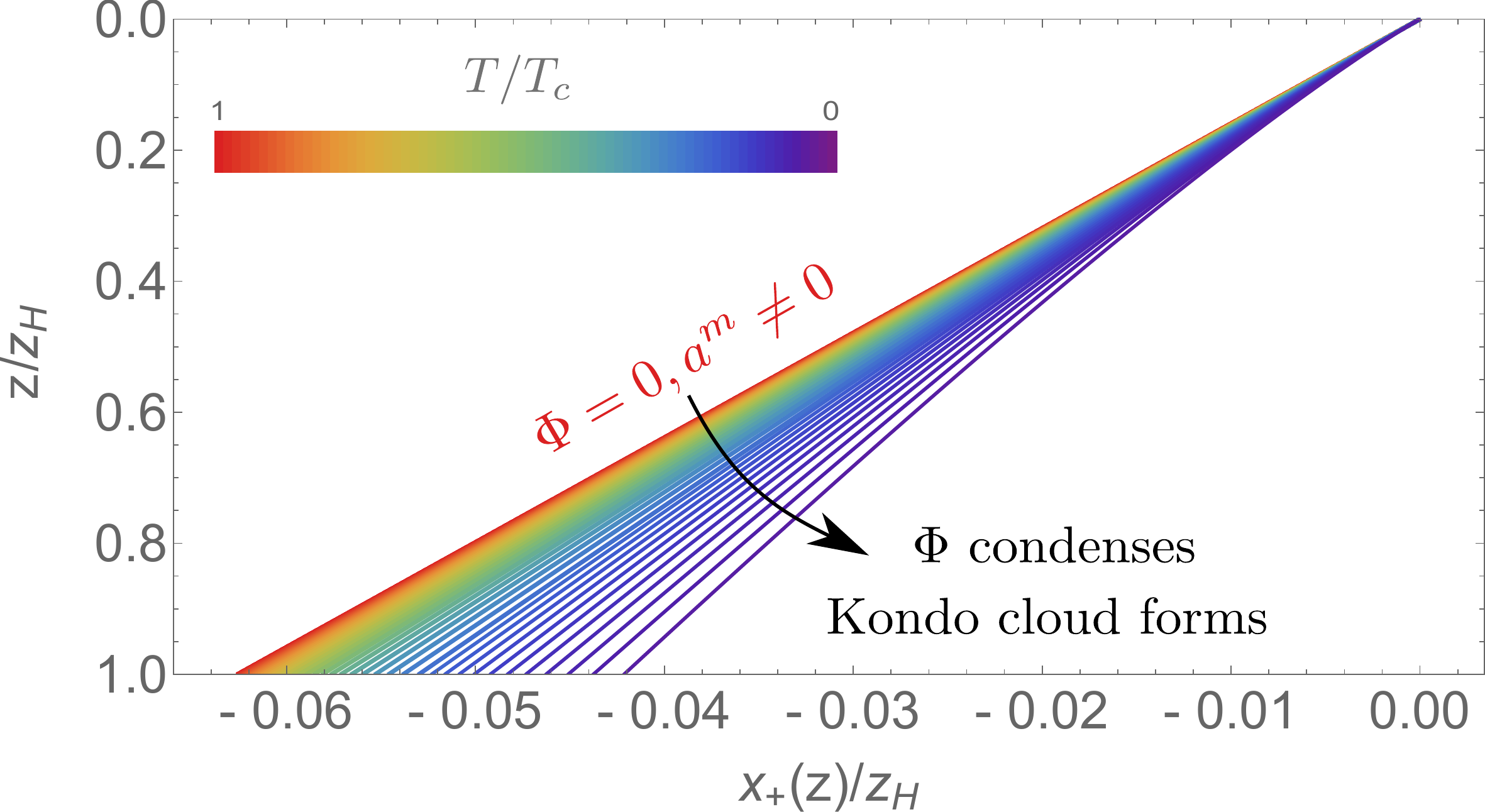}
\caption{\small \em Profile of the defect for a fixed representation of the impurity spin and different temperatures $T\leq T_c$. As $\vev{\cO}$ increases for lower temperatures, the profile moves to the right (Figure from \cite{Erdmenger:2015spo}).}\label{fig:bend}
\end{center}
\end{figure}

The matter fields on the bulk $1+1$ defect have an effect in the surrounding geometry. Since in $2+1$ dimensions gravity has no propagating degrees of freedom, the only effect is to impose some gluing condition between the geometry to the `left' and to the `right' of the defect. The conditions are the Israel junction conditions with the energy-momentum tensor of the fields on the defect (see \cite{Erdmenger:2014xya} for a detailed analysis). This approach was applied to the holographic impurity model in \cite{Erdmenger:2015spo}, the result is summarized in Figure~\ref{fig:bend}. The figure represents the `right' part of the geometry, with the $AdS$ boundary at the top and the black hole horizon at the bottom. The `left' part of the geometry is simply the mirror image. The geometry extends to the right of the figure and ends at the lines that are drawn extending to the left, so the left part of the figure is cut and glued with the geometry at the other side of the defect. The point where the lines converge at the boundary is the position of the impurity. The leftmost line (in red) is the profile of the defect in the normal phase. The slope depends on the value of the gauge field $a_t$ on the defect, so it depends on $Q$ and therefore on the spin representation. As the temperature is lowered and we transition to the condensed phase, the defect profile moves to the right as the vev $\vev{\cO}$ increases. Then, from a geometrical point of view, the effect of screening is seen in the holographic model as a reduction of the geometry.

\section{Entanglement entropy and $g$-theorem}\label{sec:entang}

The impurity affects to the state of the system by inducing a formation of the screening cloud. The entanglement entropy captures this change in the state. One can define an impurity entropy as the difference between the entanglement entropy in the presence of the impurity minus its value when the impurity is absent \cite{1742-5468-2007-01-L01001,1751-8121-42-50-504009,1742-5468-2007-08-P08003,
PhysRevB.84.041107}. For an interval of size $2\ell$ centered around the impurity field theory calculations give \cite{1742-5468-2007-08-P08003}
\begin{equation}\label{FTsimp}
\begin{split}
&S_{\rm imp}(\ell)=\frac{\pi^2 c}{6}\frac{T}{T_K} \coth\left( 2\pi\frac{\ell}{\xi_K}\frac{T}{T_K}\right), \\ 
& \ \ \ \ T/T_K, \,\xi_K/\ell \ll 1.
\end{split}
\end{equation}
Where $\xi_K\propto 1/T_K$ is identified as the screening length of the Kondo cloud \cite{PhysRevB.84.041107} and $c$ is the central charge of the CFT.  For very large sizes of the interval one can interpret $S_{\rm imp}$ as the contribution of the impurity to the thermodynamic entropy $S_{\rm imp}(\ell\to \infty)=\ln g$. This thermodynamic entropy decreases at lower temperatures, which is compliant with the $g$-theorem \cite{Affleck:1991tk,Yamaguchi:2002pa,Friedan:2003yc,Takayanagi:2011zk}
\begin{equation}\label{gthe}
T\frac{\partial \ln g}{\partial T}\geq 0.
\end{equation}
A field theory example for an RG flow satisfying the $g$-theorem may be found in 
\cite{PhysRevA.74.050305}. For further examples within holography, see
\cite{Janik:2015oja,Takayanagi:2011zk,Fujita:2011fp}.

\begin{figure}
\begin{center}
\includegraphics[width=8cm]{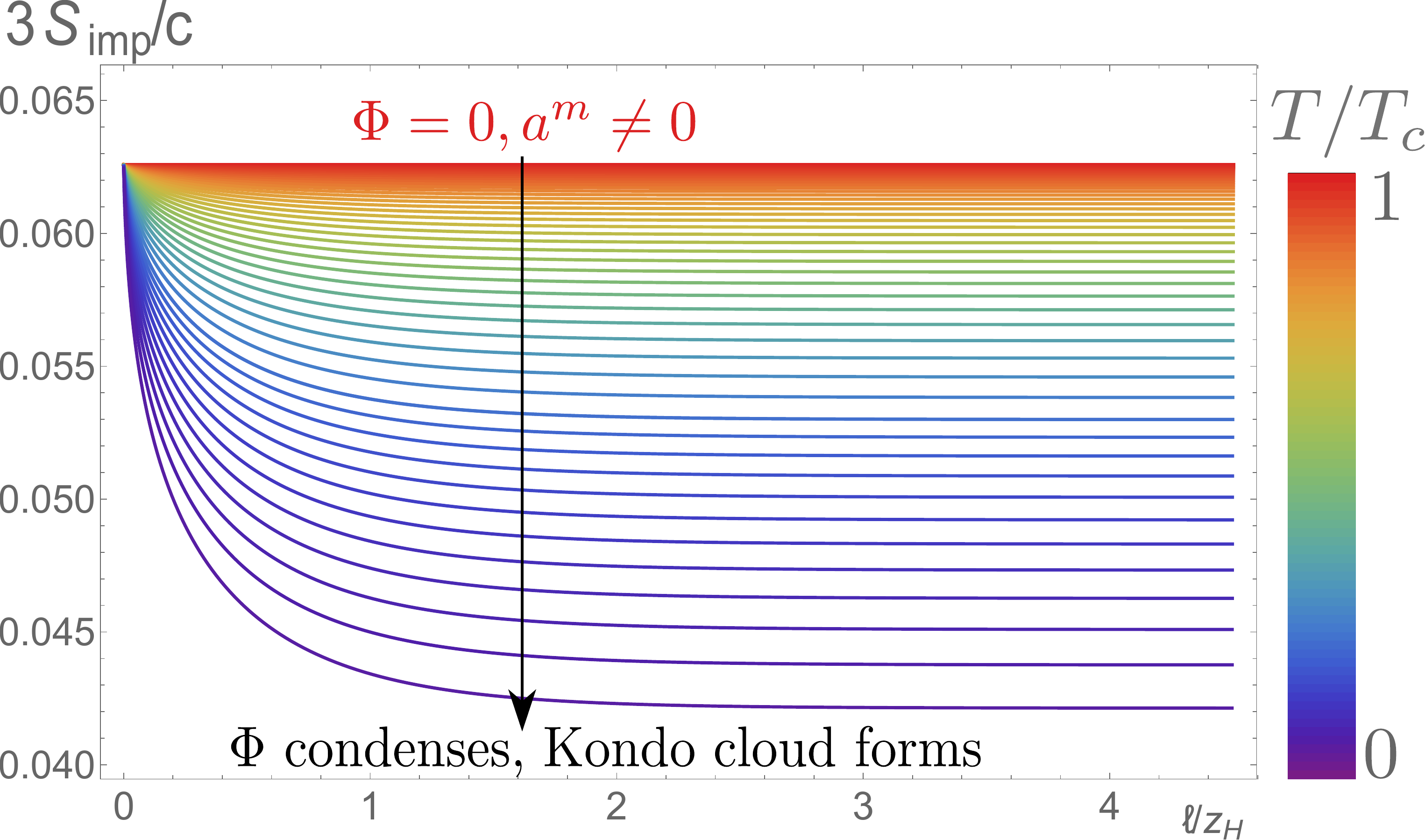}
\caption{\small \em Numerical values of the impurity entropy as a function of the size of the interval $\ell$. different lines correspond to different temperatures $T\leq T_c$. As $\vev{\cO}$ increases for lower temperatures, the impurity entropy decreases. When $\ell \to \infty$, the impurity entropy asymptotes a constant (temperature dependent) value (Figure from \cite{Erdmenger:2015spo}).}\label{fig:simp}
\end{center}
\end{figure}

In the holographic model the entanglement entropy can be computed using the Ryu-Takayanagi prescription \cite{Ryu:2006bv,Ryu:2006ef}. The entanglement entropy of an interval is proportional to the length ${\cal L}$ of a bulk geodesic connecting the endpoints of the interval at the $AdS$ boundary.
\begin{equation}
S_{EE}=\frac{c}{6}\frac{{\cal L}}{L},
\end{equation}
where $c=3 L/2 G_N$ is the central charge of the dual CFT computed according to the Brown-Hennaux formula \cite{brown1986}. In a BTZ black hole with no defects, the holographic entanglement entropy agrees with its value for a CFT at finite temperature
\begin{equation}
S_{BH}(\ell)=\frac{c}{3}\left(\frac{1}{\pi T \epsilon} \sinh(2\pi T\ell)\right),
\end{equation}
where $\epsilon\ll \ell$ is an UV cutoff. Note that the defect affects the geometry, so the impurity entropy can be defined in terms of the difference between the geodesic length in the presence and absence of the defect. The result is represented in Figure~\ref{fig:simp}. 

The upper horizontal line (in red) is the value of the impurity entropy in the normal phase. Note that it is independent of the size of the interval $S_{\rm imp}(\ell)=S_{0}(Q)$, this is consistent with taking the limit $\xi_K\to 0$ (for fixed $T_K$) in \eqref{FTsimp}. The geometric reason is that the position of the defect is found by displacing each point the same geodesic distance from the $Q=0$ profile (a vertical straight line).

In the condensed phase where the screening cloud is forming, the impurity entropy decreases as the vev of the scalar operator $\vev{\cO}$ increases. This is explained geometrically by Figure~\ref{fig:bend}. Since the position of the defect in the bulk moves to the right as the vev increases, the total length of the geodesics used to compute entanglement entropy will decrease.

\begin{figure}
\begin{center}
\includegraphics[width=6cm]{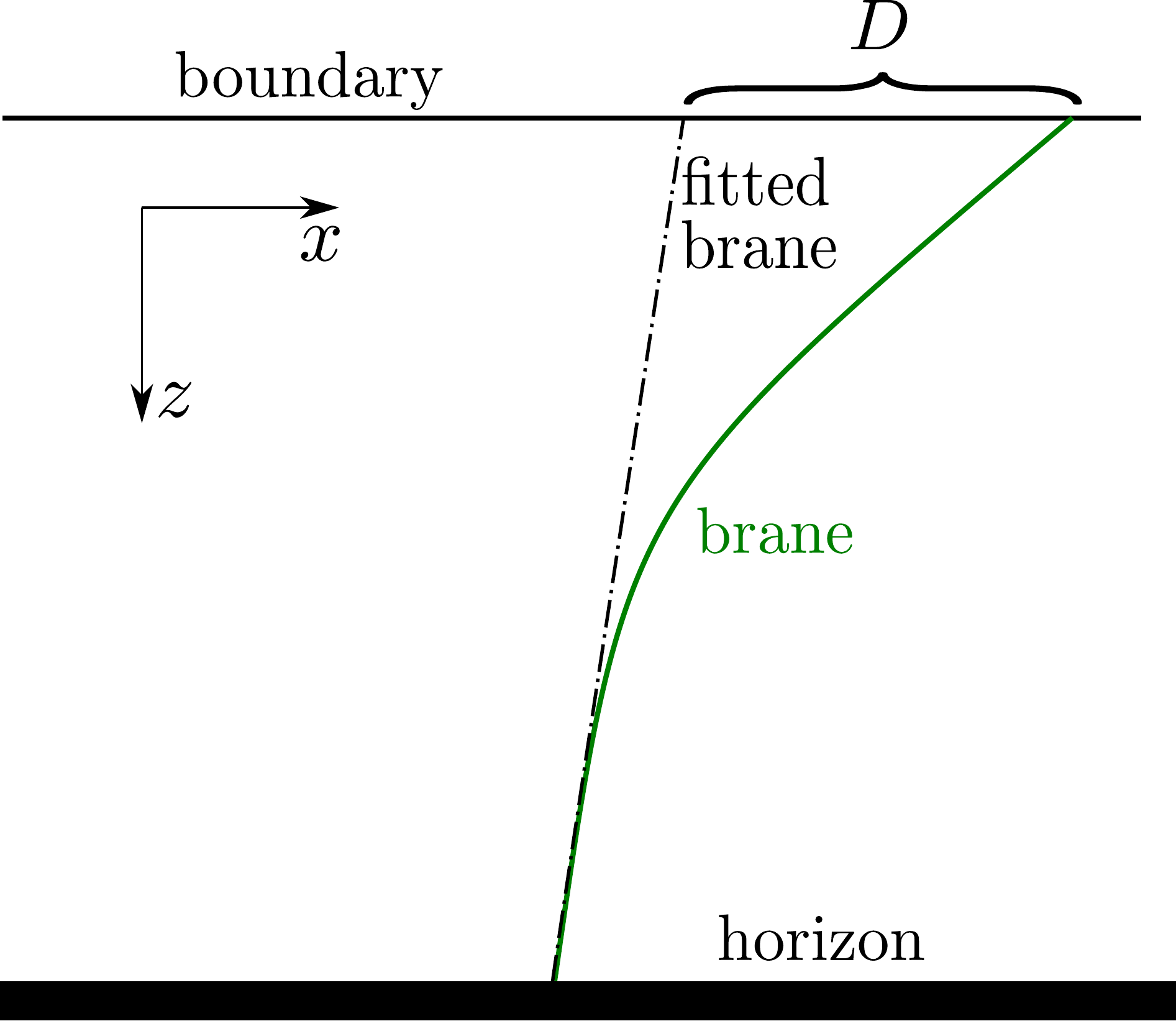}
\caption{\small \em At low enough temperatures $DT\ll 1$, the profile of the brane close to the horizon can be approximated by a normal phase profile, but displaced a distance $D$ with respect to the original position of the impurity at the boundary. The change on the slope signifies an effective change in the representation of the impurity spin, consistent with screening (Figure from \cite{Erdmenger:2015spo}).}\label{fig:Dfit}
\end{center}
\end{figure}

For low enough temperature the profile of the defect is as shown in Figure~\ref{fig:Dfit}. In the region close to the horizon, the profile can be fitted by a profile of the normal phase, but for a smaller spin representation $Q'<Q$, and displaced with respect to the original position of the impurity at the boundary by a distance $D$. For values of $\ell$ that are large enough with respect to $D$, the geodesic that determines the entanglement entropy will cross the defect in the region close to the horizon. In this case the impurity entropy can be approximated as
\begin{equation}
\begin{split}
S_{\rm imp} \simeq & S_{BH}(\ell+D)-S_{BH}(\ell)+S_0(Q')\\ 
&\simeq D\partial_\ell S_{BH}(\ell)+S_0(Q').
\end{split}
\end{equation}

\begin{figure}
\begin{center}
\includegraphics[width=8cm]{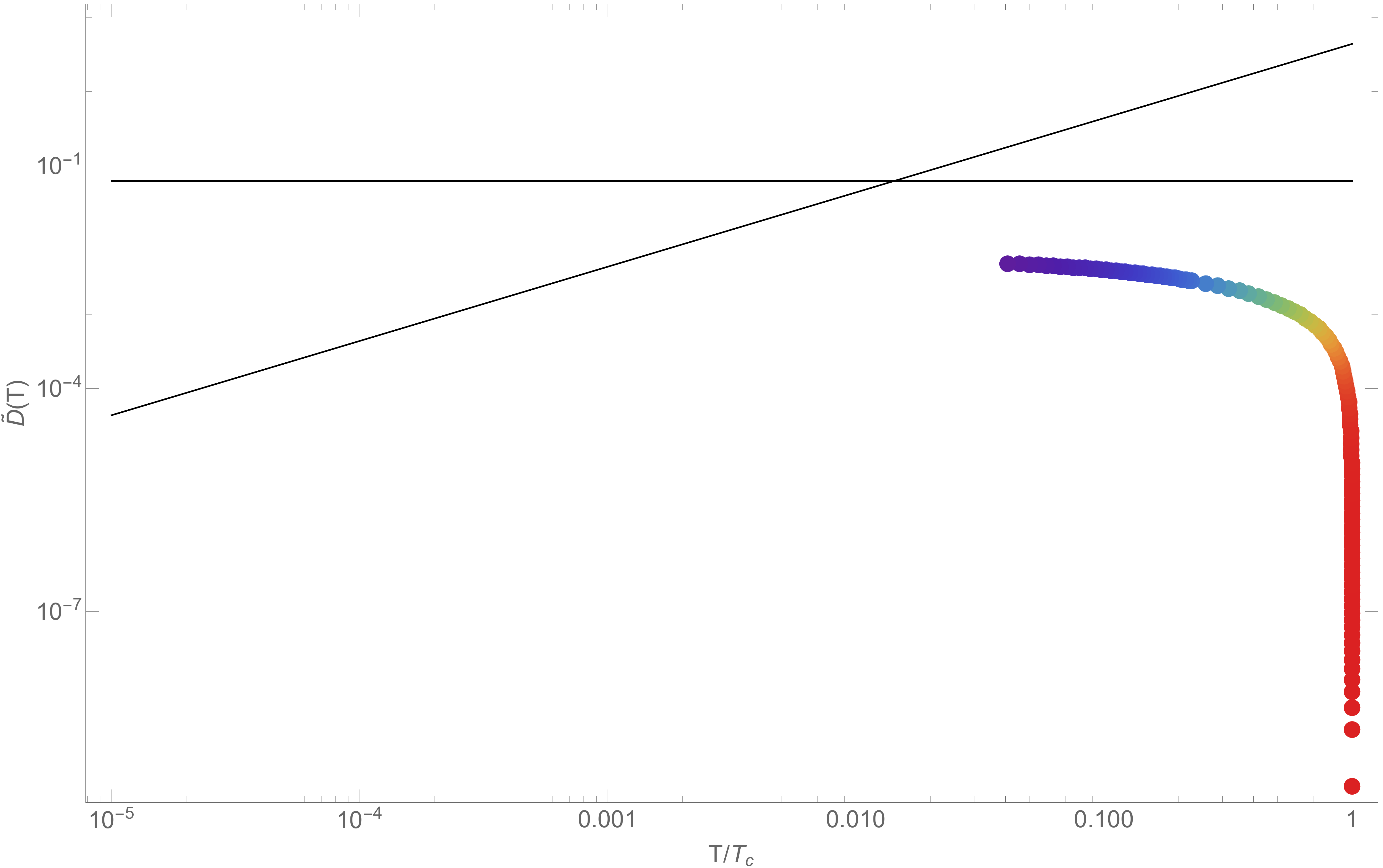}
\caption{\small\em Numerical value of $\tilde{D}=2\pi TD$ as a function of the temperature. The horizontal line is an upper bound on the value of $\tilde{D}$ in the model. The diagonal line corresponds to the field theory result at low temperatures $\tilde{D}=\frac{\pi}{2}\frac{T}{T_K}$. Note that the temperatures that are reached numerically are not low enough to check if the holographic model and an impurity coupled to a LFL have the same behavior at the IR fixed point.}\label{fig:DT}
\end{center}
\end{figure}

 This gives the analytic formula
\begin{equation}
\begin{split}
&S_{\rm imp}(\ell)\simeq S_0(Q')+\frac{2\pi c}{3}TD \coth\left( 2\pi T\ell\right), \\
& \ \ \ \  \ell T\gg 1, \ DT \ll 1.
\end{split}
\end{equation}
Comparing with \eqref{FTsimp}, we see it takes a similar form if we identify the Kondo screening length with $D=\frac{\pi}{4}\xi_K$. Note however that $D$ is a function of $T$, and the temperature dependence of the impurity entropy in the holographic model is in general different than the one where the impurity couples to a LFL. The numerical value of $\tilde{D}=2\pi TD$ is represented in Figure~\ref{fig:DT}.

\begin{figure}
\begin{center}
\includegraphics[width=8cm]{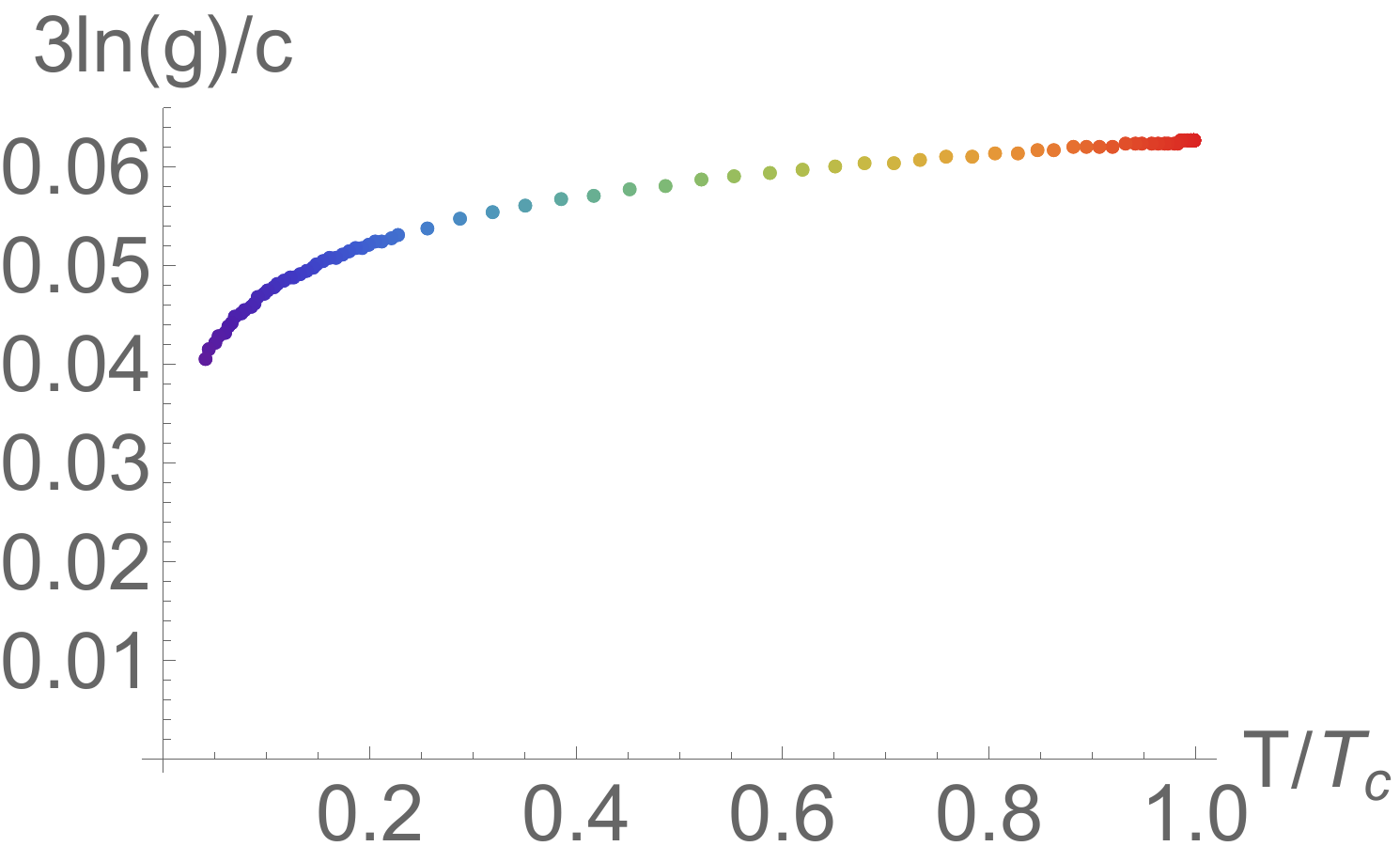}
\caption{\small \em Thermal impurity entropy as defined from $\ln g=S_{\rm imp}(\ell\to \infty)$. The entropy is monotonically increasing with the temperature, in agreement with the $g$-theorem (Figure form \cite{Erdmenger:2015spo}).}\label{fig:lng}
\end{center}
\end{figure}

Note that the value of the impurity entropy in Figure~\ref{fig:simp} asymptotes a constant when $\ell\to \infty$. We can extract the value of the thermodynamic impurity entropy $\ln g$ from this asymptotic value, we represent it in Figure~\ref{fig:lng}. Clearly, the condition \eqref{gthe} is satisfied. This is due to the holographic $g$-theorem\cite{Takayanagi:2011zk,Yamaguchi:2002pa} that links the condition \eqref{gthe} to the null energy condition (NEC) for the fields on the defect. In the model the NEC is satisfied by any configuration of matter.

\section{Conclusions and Outlook}\label{sec:out}

The results we have presented for the holographic model of a magnetic impurity suggest that the main features of the Kondo effect remain qualitatively unchanged even if the impurity interacts with a strongly coupled system, at least in the limit of large spin. The screening of the impurity has a nice geometrical realization in terms of the profile of a defect in the dual theory and an effective screening length can be obtained from the holographic calculation of the entanglement entropy. Also from the entanglement entropy one can show that the $g$-theorem is naturally satisfied in the holographic model. This seems then to be a viable model to reproduce qualitative features of more complicated situations involving magnetic impurities.

An interesting direction in the single-impurity model would be to study the effect of quenching in the Kondo coupling, which in condensed matter systems can be experimentally realized through the absorption of photons \cite{PhysRevLett.106.107402,kondoquench1}. For more than one impurity, it would be interesting to extend the results of the two-impurity model  \cite{O'Bannon:2015gwa} to a lattice of impurities and maybe reproduce some of the qualitative features of the phase diagram of heavy fermion materials.

Other possible directions would be construct holographic models dual to theories where the spin is not gauged or matter is only in a vector-like representation of the spin group, for instance in higher spin theories \cite{Klebanov:2002ja}. 




\bibliographystyle{prop2015}
\bibliography{refs}


\end{document}